\documentstyle[12pt,psfig,aasms4]{article}

\lefthead{Rosenberg A., et~al.}
\righthead{Pal~1: Another young Galactic halo globular cluster?}

\begin{document}

\title{Pal~1: Another young Galactic halo globular cluster?
\footnote{Based on observations made with the Isaac Newton 
Telescope operated on the island of La Palma by the Isaac Newton Group
in the Spanish Observatorio del Roque de los Muchachos of the
Instituto de Astrof\'{\i}sica de Canarias.}}
\author{A. Rosenberg}
\affil{Telescopio Nazionale Galileo, Osservatorio Astronomico di Padova, Italy}
\author{I. Saviane, G. Piotto}
\affil{Dipartimento di Astronomia, Universit\`a di Padova, Italy}
\author{A. Aparicio}
\affil{Instituto de Astrof\'{\i}sica de Canarias, Spain}
\author{S. R. Zaggia}
\affil{Osservatorio Astronomico di Capodimonte, Italy}

\begin{abstract}
Deep V and I CCD images of the loosely populated Galactic globular
cluster Pal~1 and the surrounding field have been obtained with the
Isaac Newton Telescope. A color magnitude diagram down to
$V_{lim}\sim24$, and a luminosity function complete down to
$V_{lim}=23.25$ have been constructed.

Adopting a reddening E(V-I)$=0.20\pm0.04$ and comparing the CMD of
Pal~1 with 47 Tucanae, we obtain a distance modulus
$(m-M)_0=15.25\pm0.25$, indicating that Pal~1 is located $3.7\pm0.4$
kpc above the Galactic disc and $17.3\pm1.6$ kpc from the Galactic
center.

Comparison with 47~Tuc and M71 shows that Pal~1 must be significantly
younger. The best fitting isochrones (Bertelli et~al, 1994) give an
age between 6.3 and 8 Gyrs, which would make Pal~1 the youngest
Galactic globular cluster so far identified, casting some doubt on the
real nature of this object.
 
The luminosity function shows some evidence for mass segregation,
consistent with the very short relaxation time. The global mass
function can be fitted with a power law of slope $x=-1.4\pm0.7$. This
mass function is anomalously flat, suggesting either a strong
dynamical evolution or an initial mass function significantly
different from most of the other halo globular clusters.

A discussion on the possible nature of Pal~1 is presented.

\end{abstract}

\section{Introduction\label{sec:intro}}

The globular cluster (GC) Pal~1=C0325+794 ($\alpha_{\rm
2000}=3^h\,33\fm4$, $\delta_{\rm 2000}=+79^\circ\,35'$; l$=130\fdg1$,
b$=19\fdg0$) is a very faint star cluster (M$_V=-2.54$; Webbink 1985,
W85) discovered by Abell (1955) on the Palomar Sky Survey plates.
From its position in the Galactic halo at about 14~kpc from the Sun,
19 kpc from the Galactic center, and 4.5 kpc above the Galactic plane
(W85), Pal 1 has been classified as a GC.

It was searched for variable stars by Kinman \& Rosino (1962, KR62).
They counted 100 stars, mostly between V=20 and V=22, but no variables
were found. KR62 also remark that, since many galaxies are seen in the
field, the absorption cannot be high.

Previous studies of the color magnitude diagram (CMD) include Ortolani
\& Rosino (1985, OR85) and Borissova \& Spassova (1995, BS95). OR85
noted that:\\ (1) The CMD of Pal~1 is peculiar for a GC. The
turn--off (TO) is well defined but the red giant branch (RGB) is very
poor, truncated at V$\sim16.3$, and the horizontal branch (HB) is
absent; \\ (2) There is some resemblance with other GCs such as
Pal~13, Pal~12, or E3. However, OR85 also suggest that ``some
resemblance to an old open cluster must be noted, mostly for the
anomalous inclination of the sub giant branch (SGB)''. \\ BS95,
reported CCD photometry in the Thuan-Gunn system of a field of
$7\times4.7$ arcmin$^2$ centered on the core of the cluster, reaching
a limiting magnitude of $g = 21.5$. They also estimate an age in the
interval 12-14 Gyr, which would make Pal 1 a typical GC.

The first determination of the metallicity was made by W85, from the
correlation between the dereddened subgiant colors and the
high-dispersion spectroscopic metallicities, obtaining [Fe/H]$=-1.01$,
adopting (B--V)$_{o,g}=1.08$ and E(B--V)=0.15 from an unpublished paper
by Da Costa (1984). The error on the estimate of the (B-V)$_{0,g}$ is
unknown (cf. Section~\ref{sec:params}) making this result very
uncertain. BS95 give [Fe/H]$=-0.79$ (No error is quoted).

With respect to the reddening, W85 gives E(B--V)=0.15 (based on the
neutral hydrogen column density plus galaxy counts), OR85 estimate
 E(B--V)$=0.1$ from the Galactic position, and BS95 give ${\rm
E(g-r)}=0.16$ (from comparison with the theoretical isochrones by Bell
\& VandenBerg 1987). The estimated distance modulus ranges between
15.68 (W85) and $15.38\pm0.15$ (BS95). BS95 derived an age of 12--14
Gyrs with a nominal uncertainty of 2 Gyrs.

The results from these works are rather uncertain, due to the small
number of cluster stars and the contamination by Galactic foreground
and/or background objects.

Here we present a new, extensive study of Pal~1, based on a set of
images collected at the INT 2.5m Telescope covering the entire cluster
(Figure~\ref{fig:image}), and a field at $10$ arcmin from the center,
taken for the purpose of studying the non-cluster star contamination.
The observations and data reductions are presented in
Section~\ref{sec:obs}. The color magnitude diagram is discussed in
Section~\ref{sec:cmd}. In Section~\ref{sec:params} we will address in
more detail the problem of the real nature of this cluster, trying to
establish its distance and age. We also present the luminosity (LF)
and mass (MF) functions (Section~\ref{sec:lf}), and the structural
parameters of the cluster (Section~\ref{sec:struc}).

\section{Observations and reductions\label{sec:obs}}

The observational data base consists of 17 frames, 9 in the V band
(with a total integration time of 7260 s), and 8 in the I band (with a
total integration time of 3230 s) centered on the cluster. We also
collected 6 V (total exposure of 6060 s) and 6 I (2490 s) frames of an
external field located $10'$ West from the cluster center with a
suitable overlap ($\simeq 1'$) with the central frame for the internal
comparison of the photometry of the two fields (see
Table~\ref{tab:obs} for a detailed log of the observations).

The data was obtained at the INT 2.5m telescope on November 8, 9 and
10, 1994, at the Observatory of Roque de los Muchachos in La Palma,
Canary Islands, Spain. The CCD used was an EEV $1242\times1152$. The
CCD scale is $0\farcs54$ per pixel, and the resulting field of view is
11.2$\times$10.3 arcmin$^2$.

The pre--processing of the images was done in a standard way, briefly
summarized below. First the constancy of the bias level across the
image was checked by taking several bias images on each night: no
gradients were detected, and we verified that the bias level was fully
consistent with the value in the overscan area of the scientific
frames. This constant value was therefore subtracted from each image.
As a second step, each image was corrected for the varying sensitivity
across the CCD frames. About seven sky flat--fields were taken per
night in both filters, and the raw images were divided by the
resulting median flats, using the appropriate reference flat according
to the night and filter.

The seeing conditions were almost constant during the three nights, so
long exposure images were averaged to obtain reference frames for each
field/filter: for the central field, we used the 8~V and 7~I long
exposures, while for the external field the 5~V and 4~I long exposures
were used. The photometry of the master frames was obtained using
DAOPHOT~II (Stetson 1987).

Calibration of the raw photometry was accomplished in three stages.
First, both the standard star magnitudes and the object frame
magnitudes were referred to a common instrumental system, ensuring
that the same fraction of the stellar flux had been sampled in both
cases. We assumed that, when a suitably large aperture is chosen, the
different percentages of the flux which are left out add negligible
corrections to the magnitudes of the standard and object stars: all
magnitudes were corrected to this common aperture. The radius was
fixed using the standard stars observed in the worst seeing
conditions. We established the limit at which the wings of the PSF
merged into the background, and set the fixed aperture $\sim 1''$
larger than this limit (i.e. 6\farcs5). For the standard stars, this
large aperture was used, while for the Pal~1 stars aperture
corrections were calculated. Actually, for the cluster frames we have
PSF magnitudes, though the instrumental PSF magnitudes ($m_{\rm PSF}$)
and the aperture magnitudes ($m_{\rm AP}$) have the same zero point
(Stetson 1987; see also Aparicio \& Gallart 1995). The aperture
corrections were calculated using the standard stars. We followed the
same method used by Aparicio \& Gallart (1995). This method is based
on the fact that the aperture corrections are a function of the seeing
alone. Figure~\ref{correct} shows the dependence of the aperture
correction on the full-width at half-maximum (FWHM) of the stellar
images. The internal dispersion (standard deviation) is of 0.02 mag.
As a second step, the dependence on the airmass and the exposure time
was removed. According to the La Palma User Manual, the shutter delay
time is of $\sim$4 milliseconds. As the exposure times on the standard
stars are between 3 and 10 sec, we expect an error $<0.001$ magnitudes
due to the shutter.

The conversion from instrumental to standard magnitudes was based on a
total of 102 measurements from multiple exposures of 23 standard stars
from the list of Landolt (1992). Independent calibrations for each
night were made. As the coefficients of the calibration curves were
the same within the errors, we adopted their mean value. Also the
atmospheric extinction coefficients were independently determined for
each night. Because of the stability of the atmospherical conditions,
we decided to adopt mean values. The (V--I) colors of the standards
range from $-0.53$ to 1.95 mag, and the V magnitudes from 12.2 to 16.1
mag. The transformation from the instrumental magnitudes to the
Landolt (1992) magnitude system was:
\begin{displaymath}
V-v = 24.95 + 0.05 (V-I), (\sigma=0.02)
\end{displaymath}
\begin{displaymath}
I-i = 24.01 - 0.02 (V-I), (\sigma=0.03) 
\end{displaymath}
where V and I are the standard magnitudes, and $v$ and $i$ the
previously defined normalized magnitudes.

Stars brighter than V=16.0 were saturated in the long exposure images.
Their magnitudes were measured on the short exposure images. After the
calibration to the standard photometric system, we checked the zero
points of the magnitudes from the short exposure images (V$\leq16.0$)
and the long exposure images (V$ > 16.0$): the mean difference is of
0.002 magnitudes in V and 0.004 magnitudes in I.

\section{The color--magnitude diagram\label{sec:cmd}}

The CMDs for the inner $30''$ and $80''$ from the center of the
cluster, and for the whole central and external frames are shown in
Figure~\ref{cmds}.

The main features of the CMD can be clearly identified in the two
central regions, though some contamination by foreground objects is
present. The CMD of the inner $80''$ is better shown in
Figure~\ref{intcmd}, where the adopted fiducial points and main
sequence (MS) widths (1 sigma) are also plotted. The fiducial points
have been obtained assuming that the color distribution is Gaussian
and determining the mean color and sigma in each 0.5 V magnitude
interval and using a k-sigma clipping algorithm (k=2.5). The fiducial
points above the TO have been obtained simply drawing a line through
the data. From the CMD in Figure~\ref{intcmd} and from the fiducial
points, we estimate that the TO of Pal~1 is located at
$V_{TO}=19.25\pm0.20$ and $(V-I)_{TO}=0.71\pm0.03$.

Given the high field contamination, one could question if the adopted
interpolation realistically represents the cluster giant branch for
$V<V_{TO}$. To test this point, we calculated the expected number of
field stars, scaling for the sampled area, on the basis of the
background field CMD: in a circle of $80''$ radius, for $V<19.25$, 4.3
field stars are expected, while in the CMD of Figure~\ref{scatole}
there are 28 stars. This means that at least 80~\% of the stars from
$V_{TO}$ to $V\sim16$ are likely to be real cluster members. We
calculated also the expected number of field stars in square boxes of
arbitrary size within the CMD, again after scaling for the sampled
area: the result is sketched in Figure~\ref{scatole}, where the stars
in the inner $80''$ are plotted and the numbers represent the field
stars expected in each box. It is quite evident that most of the
``evolved'' stars in Figure~\ref{intcmd} and Figure~\ref{scatole} are
probably cluster members.

The relative number of stars in different regions of the CMD may also
be estimated from stellar evolution theory. Using the isochrones by
Bertelli et~al, (1994, B94) and assuming a Salpeter MF (the result
does not change significantly adopting a different MF slope), we can
calculate the number of stars which are expected in each phase of the
CMD for different metallicities and ages. As reported in
Table~\ref{tab:BEF}, the counts within $80''$ from the center are
consistent with the expected ones, which have been calculated assuming
ages from $6.3$ to $20$ Gyrs and metallicities from $-0.3$ to $-1.0$.
The star counts have been normalized to the total number of stars.
Furthermore, four RGB stars of Pal~1 have been studied
spectroscopically (Rosenberg et~al, 1997), and their radial velocities
are compatible with membership in the cluster.

One could ask if any of the 8 stars in the box of the CMD defined by
$0.65<$(V--I)$<0.85$ and $18.75<$V$<19.25$ are binaries.
Unfortunately, the broadening of the MS due to the photometric errors
and the small statistical sample does not allow us to draw any
conclusions about the presence of a binary sequence.  The presence of
binaries around the TO could affect the magnitude (but not
significantly the color) of the TO itself and the discussion of the
cluster age.  We can make an estimate of the possible error in the
age by assuming that 50 per cent of the stars are binary objects. In
this case, the true TO luminosity would be $\sim 0.35$~mag fainter.
Since the age will be estimated from the color difference between the
TO and the RGB (cf. Sect.~\ref{sec:params}), a change in color of the
RGB reference point due to the change in the TO location, would be
0.03~mag towards the blue. This would increase our measured age by
$\simeq 2$~Gyrs.  On the other side, making the TO 0.35 mag fainter,
would increase the age by at most $\sim 3$~Gyr (from eq. 5b in
Buonanno, Corsi, \& Fusi Pecci 1989).

A final comment on the possible presence of HB stars. In their
preliminary work, OR85 could not find any star in this phase, while
BS95 suggest that 5 stars located $\simeq3.2$ mag above the TO in
their CMD could be HB stars. Statistically $\sim1$ star is expected
to be in this phase accordind to the evolutionary tracks, so we can
not exclude the possibility that one or two stars at V$_{\rm
HB}\sim16.3$ in Figure~\ref{intcmd} are part of the red clump of the
cluster's HB. It is impossible to say anything on the presence of a
HB in the complete CMD of Figure~\ref{cmds}.

From this discussion, it must be clear that any cluster parameter in
the literature (and in the following) derived from the location of the
HB, must be considered only tentative and uncertain.

\section{Determination of the main parameters\label{sec:params}}

Since a precise value of the metallicity is crucial for the age
determination, we have carried out spectroscopic observations in order
to determine this parameter following the method of Armandroff \&
Da~Costa (1986,1991). The detailed analysis of our data set, which is given
in Rosenberg et~al, (1997), yields a value [Fe/H]$=-0.60\pm0.20$~dex,
and we will adopt this metallicity in the following.

As already discussed in the introduction, all the previous
determinations of the reddening established a value around E(B--V)$ =
0.15$; in order to have an independent estimate of this quantity, we
inspected the Burstein \& Heiles (1982) maps, which give an E(B--V)
between 0.12 and 0.15. In the present paper, we will assume
 E(B--V)$=0.15\pm0.03$, which implies E(V--I)$=0.20\pm0.04$ (adopting
E(V--I)$=1.28$ E(B--V), Dean et~al, 1978).

With the adopted values for the reddening and metallicity, it is now
possible to determine the distance modulus of Pal~1 by comparison with
other clusters of similar metallicity, after establishing a reliable
set of parameters for the reference clusters.

In this respect, 47~Tuc [Alcaino \& Liller 1987, (AL87), Grundahl 1996
(G96), Kaluzny et~al, 1997 (K97)] and M~71 [Richer \& Fahlman
(1988,RF88), Hodder et~al (1992,H92), Grundahl 1996 (G96)] are the
only GCs that can be used for this comparison. Their metallicity is
around [Fe/H]$ = -0.7$ and they have accurate published $VI$
photometry from the RGB to several magnitudes below the TO (Note that
the diagram by G96 is the only CMD available in V vs. (V-I) for M71).
A careful discussion of the 47~Tuc parameters can be found in Hesser et~al,
(1987, H87), and we will assume their values for the cluster
reddening, metallicity and distance: E(B--V)$ = 0.04\pm0.01$
[corresponding to E(V-I)=0.05], [Fe/H]$ = -0.8$ and $(m-M)_V =
13.35\pm0.2$. The quoted distance modulus is a mean of the two values
found by H87, namely $(m-M)_V = 13.4$ and $13.3$ from a comparison
with the theoretical ZAHB and the field subdwarfs, respectively.
In the following, we will refer to 47~Tuc as reference cluster.

We plotted the three available CMDs of 47~Tuc together, and evaluated
the zero--point differences in the turnoff region. There are overall
differences of less than 0.02~mag in color, with the K97 sequences in
between the other two. Given the high statistical significance (more
than 15,500 stars) and the depth ($\sim3.5$~mag below the TO) of the
K97 sample, we chose this dataset as the 47~Tuc template.

Figure~\ref{compare} displays the absolute fiducial sequences of
47~Tuc (open triangles) and, M71 (open circles). 
After a shift of $\Delta(V-I)=0.24$ and $\Delta(m-M)_V=0.38$
the fiducial line of M71 perfectly overlaps 47~Tuc. The sequence of
47~Tuc has been used to determine the distance modulus of Pal~1.
After correcting the CMD of Pal~1 for the adopted reddening, the
apparent distance modulus has been computed by fitting its lower MS to
that of 47~Tuc. The fit was carried out in the absolute magnitude
interval $5<M_V<7$, where evolutionary effects due to possible
differences in age should be minimal.

Allowing E(V--I) to vary from 0.16 to 0.24, the apparent distance
modulus of Pal~1 varies in the interval $15.51 <(m-M)_V< 15.94$, as
shown in Figure~\ref{compare}. Taking into account the uncertainty in
the calibration, in the adopted reddening, in the fit, and in the
distance of the comparison clusters, the absolute distance modulus is
$(m-M)_0=15.25\pm0.25$.

As shown in Figure~\ref{compare}, a good match of the
Pal~1 giant branch would be obtained for a value of the reddening
E(V--I)$ \simeq 0.18$. On the other hand, the position of the RGB is
controlled not only by a cluster metallicity, but also by its age,
and, as we will see later in this section, Pal~1 seems to be younger
by at least 6 Gyr than normal GCs. Therefore, we expect that the
stars which are evolving along the RGB of this cluster are more
massive (and hence hotter) than those of the comparison clusters
(which have a `normal' age, cf. H87 and G96).

In this case, the reddening must be higher than the value obtained
from a direct comparison with the sequences of 47Tuc and M71, since
its bluer RGB requires a smaller shift in order to be superposed on
the reference RGB. In order to estimate this effect, we analyzed a set
of isochrones from Bertelli et~al, for [Fe/H]$=-0.7$, and found a
relative shift of $0.03\div0.04$ mag between the RGBs for the expected
age interval. A value E(V--I)$ \simeq 0.21-0.22$ is therefore a better
estimate of the cluster reddening. This is almost coincident with the
adopted value.

The most striking feature of the Pal~1 CMD which emerges from the
comparison with M~71 \& 47 Tuc is the larger color difference between
the TO and the RGB base: it is well-established (see e.g. Vandenberg,
Bolte \& Stetson 1990) that the color width between the TO and the RGB
is a function of the cluster age and metallicity. Since the template
clusters and Pal~1 have similar metallicities, the discrepancy which
is observed for Pal~1 is most likely due to a younger age.

This result is confirmed by a direct comparison of the Pal~1 CMD with
a set of theoretical isochrones taken from the library of B94. In
order to put a constraint on the subset of model ages to be used, we
first compare in Figure~\ref{morfologia} the CMD of Pal~1 with the
models, registering both of them to the TO. In this way, the color
width between the TO and the RGB base can be used to select a
subsample of isochrones. The figure clearly shows that a standard
16~Gyr isochrone is not able to reproduce the morphology of the
diagram, and we have to assume an age between 6.3 to 8 Gyrs in order
to fit the cluster's fiducial points. This would make
Pal~1's age of the order of the oldest open clusters. 

This result is furtherly confirmed by Figure~\ref{fit_isocrone} which
shows that the 8~Gyr isochrone is the one that better reproduces the
overall morphology of the CMD in the $(V-I)_0$, $M_V$ plane when we
adopt the distance modulus and reddening previously discussed. A
slight blueward offset of 0.03~mag was applied to the isochrones in
order to superpose the main sequences and giant branches: this offset
is entirely within the uncertainties in the value of the reddening.

If we adopt the TO magnitude and the (very uncertain) HB magnitude
discussed in Section~\ref{sec:cmd}, we would have a $\Delta
V^{HB}_{TO}=2.95$, corresponding to an age of 8.3 Gyr (Buonanno et~al,
1989, eq. 5b).

In summary, from the comparison with the theoretical isochrones and
the old GCs we can conclude that Pal~1 must have an age significantly
lower than the bulk of the Galactic GCs.

The adopted distance modulus corresponds to a distance from the Sun of
$11.2\pm1.3$~kpc, a distance from the Galactic center $R_{\rm
GC}=17.3\pm1.6$~kpc and a height $Z_{\rm GP}=3.7\pm0.4$ above the
Galactic plane (we adopted a distance Sun--Galactic Center
$R=8.0\pm0.5$ kpc, Reid 1993)

In view of its position in the Galaxy, it is not very likely that
Pal~1 is an old open cluster, though this possibility can not be ruled out.
We note that in the compilation by Friel (1995, F95) of the oldest
open clusters, none of them has $Z_{\rm GP}>2.4$~kpc. On the other
hand, the metallicity, though very low for an open cluster, would
still be consistent with the radial abundance gradient for the old
open clusters proposed by F95. As discussed in
Section~\ref{sec:struc}, Pal~1 seems to be more concentrated than
normal old open clusters, while its concentration parameter $c=1.6$ is
typical for a globular cluster (Trager, King \& Djorgovski 1995).

\section{Luminosity and Mass functions\label{sec:lf}}

We have obtained the stellar luminosity functions for two radial
regions of Pal~1: inner, $r\leq 30''$, and outer, $30''< r <
270''$. We extracted the LFs using only stars falling within 3 sigmas
of the fiducial line of the MS. In order to facilitate the star
counts, we linearized the MS by subtracting from each star the V--I
color corresponding to the V magnitude along the MS. We obtained the
LFs in one magnitude bins, with the exception of the first bin which
includes all the evolved stars. The two LFs are reproduced in
Figure~\ref{fdlfinal}, together with the LF of the background obtained
in the same way. The outer LF of Pal~1 has been arbitrarely shifted in
order to match the first two bins of the inner LF.

Each LF was corrected for completeness. We used an approach similar
to that described in Piotto et~al (1990). The completeness was
estimated in the magnitude range V=19.5-25.0 by adding artificial
stars to the original frames in 0.25 magnitude intervals, and checking
what percentage was recovered in each region, after a reduction stage
equal to that of the original images. The artificial stars were added
to the average frames. The I magnitudes for the stars were set using
the (V--I) color of the MS. We considered as recovered stars only those
identified both in the V and I frames. For the central field we reach
the $50\%$ completeness at $V\simeq23.7$, and $V\simeq23.5$ for the
external one. When doing star counts we stopped at $V=23.25$, well
above the $50\%$ limit (see Table~\ref{tab:LF}).

The LFs were also corrected for field star contamination by
subtracting the star counts coming from the external frame ($r>350''$,
see Figure~\ref{fdlfinal}) to the counts of the central
frame, after correcting for completeness and area coverage.

The completeness corrected star counts (Columns 2,4, and 6) and the
luminosity function (Columns 3,5,and 7) for each radial bin are
reported in Table~\ref{tab:LF}, where in the last column we present
also the values of the completeness correction adopted for the central
and the external field. 

The error bars of the LFs include the Poisson errors of both the cluster
and background counts, plus the uncertainty in the completeness
correction.

The two LFs look different (though the small number of stars does not
allow to assess the statistical significance of this result),
with the internal LF flatter than the external one as can be expected as a
consequence of the equipartition of the energy. In a loosely populated
cluster like Pal~1, dynamical evolution is very fast. The relaxation
time is extremely short if we compare it with the typical values for
GCs. In particular, Pal~1 has the shortest relaxation time in the
compilation by Djorgovski (1993, D93), with a relaxation time at the
half-mass radius T$_{rh}=3.3\times 10^7$~Gyr. Therefore, we do expect to
detect mass segregation effects in its radial stellar distribution.

We calculated the MF corresponding to the LFs presented in
Figure~\ref{fdlfinal}, using B94 models and the cluster parameters
adopted in Section~\ref{sec:params}. The two annular MFs and the
global MF (i.e. the MF obtained for the entire cluster) are presented
in Figure~\ref{mffinal} (the global MF is shown shifted up by 0.2
units for clarity).

The best fitting power laws of the form $\xi=\xi_0m^{-(1+x_{\rm MF})}$
($x_{\rm MF}=1.35$ for the Salpeter MF) are: $x_{\rm MF}=-2.9\pm1.3$ for
the inner MF, and $x_{\rm MF}=-0.7\pm0.6$ for the outer one. The global
MF (obtained from the global LF) has a slope $x_{\rm MF}=-1.4\pm0.7$. 
As expected from the LFs, the MFs for the two annuli showed in Figure 10
are different (though this result is of low statistical significance), 
with the inner MF showing a reversed slope with respect to the outer one.

\section{Dynamical analysis\label{sec:struc}}

In order to shed light on the real nature (open {\it vs.} globular) of
Pal~1 we have also compared the structural parameters of Pal~1 with those
of the globular and open cluster populations.

We derived the density profile of Pal~1, making radial star counts in
equal-logarithmic steps, adopting the same selections in the CMD used
to extract the LF. The result is shown in Figure~\ref{profile} in
which we show our star counts (closed circles), corrected for
completeness and field object contamination. The open circles
represent the light profile of Pal~1 published by Trager, et~al,
(1995, T95). Our star counts, transformed into magnitudes, have been
shifted to match the T95 surface brightness profile.

One of the main problems in obtaining the profile of Pal~1 was to find
its center. Choosing a wrong center in a loosely populated cluster
with a small number of stars like Pal~1 would produce spurious trends
in the central part of the profile (T95). We tried many algorithms
like the ones listed in Picard \& Johnston (1993), finding that the
most reliable results could be obtained with the iterative centroiding
algorithm by Auri\'ere (1982).

The density profile clearly shows the presence of a flat core.
Figure~\ref{profile} shows no evidence for flattening of the T95
profile, probably consequence of incorrect centering of the star
counts or an erroneous crowding correction in their starcounts (the
most central point in T95 is from former star counts by King et~al,
1968). The small number of cluster members has prevented us from
extracting density profiles for stars of different mass: the density
profile presented here is the profile for all the stars counted to the
limit of the photometry (and corrected for completeness). Our data
set allows a reliable estimate of the field object contamination,
which is quite high (as already seen from the CMD of
Figure~\ref{cmds}). The background level was determined by counting
stars outside $350''$ from the center of the cluster: {\it i.e.}, in
the external field. The value used for the Pal~1 profile is $4.55$
stars/arcmin$^2$, as shown in Figure~\ref{profile}.

We have fitted the radial profile with a single mass isotropic King
(1966) model to extract the structural parameters of Pal~1. The best
fitting model has a concentration parameter $c=1.6$, a core radius of
$r_c=13.5''$\, and a half-mass radius $r_h=41''$, values that are well
within the range observed for other Galactic GCs. The concentration
parameter $c=1.6$ is higher than the typical open cluster
concentration parameters: none of the open clusters listed in Danilov
\& Seleznev (1994) has such a high $c$ value. The central surface
brightness is $\mu_V(0)=20.93\pm0.25$, while the surface brightness at
the half-mass radius is $\mu_V(r_h)=23.52\pm0.25$. The tidal radius
obtained from the model is $r_t=525''$ even if beyond $\simeq350''$
the star density of the model imply a total number of cluster stars
less than one\footnote{BS95 found a higher value for $r_c$\ and $r_t$\ (tidal
radius), but without a direct measure of the background/foreground
star contamination.}. This is due to the small total number of cluster stars:
integrating the global LFs presented in Table~\ref{tab:LF}, down to the
adopted limit for our photometry ($V=23.25$), we have a total number of
$N_{tot}=342\pm21$ stars. which correspond to a
total integrated absolute magnitude $M_{V}=-2.54\pm0.50$. This value
places Pal~1 at the lower end of the globular cluster luminosity
function. Using the above estimate for the total absolute magnitude
and the relation by Mandushev, Spassova, and Stanaeva (1991, eq. 4) 
we obtain log$(m/m_\odot)=3.1\pm0.3$ for the total cluster mass.

Even if the structural parameters of Pal 1 are very similar to the
ones observed in other GCs, its global MF is different. As shown in
Figure~\ref{xrz}, the Pal~1 global MF slope falls completely outside
the correlation between the mass function slope with the cluster
Galactic position ($R_{GC}$, $Z_{GP}$), and metallicity [Fe/H] proposed
by Djorgovski, Piotto, \& Capaccioli (1993, DPC). This could be an
evidence that the initial MF of Pal~1 was significantly different from
the other 21 GCs shown in Figure~\ref{xrz}. Another possibility is
that, while the correlation by DPC holds for clusters with a mass
function evolved by tidal shocks (Capaccioli, Piotto, \& Stiavelli
1993), the MF of Pal~1 could be evolved as a consequence of another
dynamical mechanism that can be tentatively identified as the stellar
evaporation from the cluster. Indeed, in the recent work on the
destruction rate of GCs by Gnedin \& Ostriker (1996), Pal~1 has one of
the highest evaporation destruction rates, while the values due to
tidal shocks are comparable to the ones for other halo GCs (Gnedin \&
Ostriker 1996). Moreover, Johnstone (1993, J93) analyzing the
evaporation escape rate of stars in GCs using multi-mass King models,
found that the MF can be modified in a remarkable way. The effect is
always toward a flattening of the MF and depends on the relaxation
time: it is faster for very low concentration and/or loosely populated
clusters. However, we cannot really say, without a simulation suited
to Pal~1, if its observed MF is a consequence of evaporation only.
Solving this problem would give us some information on the formation
of this cluster. In the case of Pal~1, we can only calculate
(following J93) the present evaporation destruction time, $T_{ev}$,
assuming a constant rate of evaporation: $T_{ev}=24\times
T_{rh}=0.74$~Gyr: this shows that Pal~1 is very likely on the verge of
destruction.

\section{Summary and Discussion \label{sec:conc}}

The principal parameters of Pal~1 obtained in the present work are
summarized in Table~\ref{tab:finalpar}. The main results are:

\begin{itemize}

\item[--] Pal~1 is very loosely populated: $N_{tot}=342\pm21$ stars down
to V mag 23.25, $M_V=-2.5\pm0.5$, and $M=1300\pm600m_\odot$

\item[--] The mean features of the CMD are clearly defined for the
internal region, though no obvious HB stars can be
identified. Consequently, any cluster parameter based on the HB
location must be considered very uncertain.

\item[--] In comparison to 47 Tucanae and M~71, 
Pal~1 seems to be younger than the typical galactic GCs.

\item[--] By comparison with isochrones, Pal~1 appears to have an age 
between 6.3 and 8 Gyr. Consequently, if we consider Pal~1 as a GC, it
would be the youngest among the Milky Way GCs and the most metal rich
halo GC. Alternatively, it would be one of the oldest open clusters.

\item[--] A surface density profile is presented and compared with
previous data. Pal~1 does not show any evidence of core collapse: its
central profile is flat. The best fitting single mass King model
gives: $\mu_V(0)=20.93$, $c=1.6$ and $r_c=13.5''$. If Pal~1 is an old
open cluster, it is very peculiar, not only because of its position in
the Galaxy ($R_{GC}=17.3$~kpc, $Z_{GP}=3.7$~kpc), but also for its
concentration parameter ($c=1.6$), which is higher than in any other
open cluster.

\item[--] The observed luminosity function 
shows some evidence of mass segregation, as expected on the basis of 
the very short relaxation time of the cluster. 

Consequently, also the MF slope varies
with radius, though this result is statistically less significant.

\item[--] The global MF has a slope $x_{MF}=-1.4\pm0.7$.
The Pal~1 MF slope does not follow the general trend with metallicity
and Galactic position proposed by DPC: if this is an indication of
evolution, the MF of Pal~1 has not been modified only by tidal shocks.
In view of the short relaxation time, it might have been modified by
evaporation. Another interesting possibility is that the initial MF
of Pal~1 was quite different from that of other halo GCs.

\item[--] Evaporation implies a destruction time of $T_{ev}=0.74$~Gyr:
Pal~1 is very likely on the verge of destruction.
\end{itemize}

In conclusion, interpreted as an open cluster, Pal~1 would be very
peculiar, at least in terms of position and morphology. On the other
hand, we point out that there is a growing body of evidence for the
existence of a small group of Galactic GCs that appear to be
significantly younger than the average Galactic GC population. Pal~12
(Stetson et~al, 1989; Gratton \& Ortolani 1988) and Rup~106 (Buonanno
et~al, 1993; Da~Costa, Armandroff, \& Norris 1992) are 3-5 Gyrs
younger than the bulk of GCs having similar metallicities. Arp~2
(Buonanno et~al, 1995a), Terzan~7 (Buonanno et~al, 1995b) and IC4499
(Ferraro et~al, 1995) are other young candidates recently discovered.
Pal~1 might be the youngest member of this group.

Many authors have tried to understand the existence of these young
globular clusters: formation from clouds that have survived in the
halo (Searle \& Zinn 1978), inter-galactic clusters captured by the
Milky Way (Lin \& Richer 1992, Buonanno et~al, 1995b), objects formed
via interactions between the Galaxy and the Magellanic Clouds (Fusi
Pecci et~al, 1995) or other satellites which may have long since
merged with our Galaxy (Majewski 1993, Lynden-Bell \& Lynden-Bell
1995). Though other explanations are possible, the fact that the MF
of Pal~1 seems to differ significantly from the MFs of 21 old GCs
might be an interesting indication of a different origin for this
object. This fact coupled with the young age might be a further
evidence that Pal~1 is to be considered a member of a different
globular cluster population, with a different formation process and
time. It would be of great interest to compare the MF of this cluster
with the MFs of other young candidates in order to check this
hypothesis. Dynamical calculation of the evaporation effects are also
needed, in order to correctly interpret the MF of Pal~1.

\acknowledgments
This project has been partially supported by the Agenzia Spaziale
Italiana. The observation run has been supported by the European
Commission through the Activity ``Acces to Large-Scale Facilities''
whithin the Programme ``Training and Mobility of Researchers'',
awarded to the Instituto de Astrofisica de Canarias to fund European
Astronomers access to its Roque de Los Muchachos and Teide
Observatories (European Northern Observatory), in the Canary Islands.
We recognize partial support by the Instituto de Astrofisica de
Canarias (grant P3/94) and by a Spanish-Italian integrated action. We
thank Prof. Jack Sulentic for the careful reading of the manuscript.

\newpage

\figcaption[rosenberg.fig1.ps]{Central 4.3$\times$4.3 arcmin image of
Pal~1, taken with the I filter and an exposure time of 600 s.
\label{fig:image} }

\figcaption[rosenberg.fig2.ps]{
The aperture photometry correction as a function of the FWHM for all the
observed standard stars. The dotted line represents a linear fit to
the data. 
\label{correct}}

\figcaption[rosenberg.fig3.ps]{Color-magnitude diagrams for three Pal~1 
regions and for the external field. 
\label{cmds}}

\figcaption[rosenberg.fig4.ps]{Fiducial points for Pal~1 plotted over
the CMD for the central $80''$. The error bars represent the
$1\sigma$ width of the MS. 
\label{intcmd}}

\figcaption[rosenberg.fig5.ps]{Color-magnitude diagram for the central
$80''$: each box contains the expected number of field stars within
the sketched color and magnitude limits. 
\label{scatole}}

\figcaption[rosenberg.fig6.ps]{Determination of the Pal~1 distance modulus from
the comparison of its fiducial line with the fiducial lines of 47 Tuc. 
The fiducial line of M71 is also shown. (See text for details). 
\label{compare}}

\figcaption[rosenberg.fig7.ps]{The Pal~1 fiducial points and a set of isochrones from
Bertelli et~al, (1994) has been registered so that the TOs
coincide. The metallicity of the isochrones is $Z = 0.004$ and the
ages are, from left to right, 16, 12.5, 10, 8 and 6.3 Gyrs; this
comparison shows that the shapes of the Pal~1 CMD is compatible with
an age in the interval 6.3 to 8 Gyrs. 
\label{morfologia} }

\figcaption[rosenberg.fig8.ps] {A subset of isochrones from Bertelli et~al, 
(1994) overplotted on the Pal~1 CMD; the diagram was corrected for the 
assumed distance modulus and reddening (see text), and the isochrones 
were shifted by 0.03~mag blueward to match the observed MS. An age of 
8~Gyr is the one that better reproduces the data. 
\label{fit_isocrone} }

\figcaption[rosenberg.fig9.ps]{ Observed luminosity functions for the
cluster center ($r<30''$), the outer part ($30''<r<270''$), and for
the background field outside $r>350''$ from the center. The cluster
LFs were corrected for completeness and field contamination; the outer
LF was normalized to the inner one using the first two bins 
(i.e. shifted up by 1.853 units).
\label{fdlfinal}}

\figcaption[rosenberg.fig10.ps]{Mass functions of Pal~1 obtained from
the LFs of Figure~\protect{\ref{fdlfinal}} using B94 models for 
$[Fe/H]=-0.7$ and age=8.0~Gyr.
\label{mffinal}}

\figcaption[rosenberg.fig11.ps]{Close dots represents the radial star
counts coming from the present photometry of Pal~1. Open circles are
the data from Trager et~al, (1995), while the solid line represents the
best fitting isotropic single mass King model. See text for a detailed
explanation. 
\label{profile}}

\figcaption[rosenberg.fig12.ps]{Trivariate relation between the position 
($R_{GC},Z_{GP}$), metallicity [Fe/H] and the slope of the global mass
functions for 21 old GCs of our Galaxy (close dots). The star
represents the position of Pal~1. 
\label{xrz}}

\dummytable\label{tab:obs}
\dummytable\label{tab:BEF}
\dummytable\label{tab:LF}
\dummytable\label{tab:finalpar}


\newpage

\begin{thebibliography}{}

\bibitem{} Abell, G. O. 1955, \pasp, 67, 258.
\bibitem{} Alcaino, G., \& Liller, W. 1987, \apj, 319, 304.
\bibitem{} Aparicio, A., \& Gallart, C. 1995, \aj, 110, 2105.
\bibitem{} Armandroff, T. E., \& Da Costa G.S 1986, \aj, 92, 777.
\bibitem{} Armandroff, T. E., \& Da Costa G.S 1991, \aj, 101, 1329.
\bibitem{} Auri\'ere, M. 1982, \aap, 109, 301.
\bibitem{} Bell, R., VandenBerg, D. A. 1987, \aaps, 63, 335.
\bibitem{} Bertelli, G., Bressan, A., Chiosi, C., Fagotto, F., \& Nasi E. 1994, 
\aaps, 106, 275
\bibitem{} Borissova, J., \& Spassova, N. 1995, \aaps, 110, 1
\bibitem{} Buonanno, R., Corsi, C. E., \& Fusi Pecci, F. 1989, \aap, 216, 80
\bibitem{} Buonanno, R., Corsi, C. E., Fusi Pecci, F., Richer H. B., \& Fahlman 
G. G. 1993, \aj, 105, 184
\bibitem{} Buonanno, R., Corsi, C. E., Fusi Pecci, F., Richer H. B., \& Fahlman 
G. G. 1995a, \aj, 109, 650
\bibitem{} Buonanno, R., Corsi, C. E., Pulone, L., Fusi Pecci, F., Richer H. B., 
\& Fahlman G. G. 1995b, \aj, 109, 663
\bibitem{} Burstein, D., \& Heiles, C. 1982, \aj, 87, 1165
\bibitem{} Capaccioli, M., Piotto, G., \& Stiavelli, M. 1993, \mnras, 261, 819
\bibitem{} Da~Costa, G. S., Armandroff, T. E., \& Norris, J. E. 1992, \aj, 104, 
154
\bibitem{} Danilov, V. M., \& Seleznev, A. F. 1994, Astron. and Astroph. Trans., 
6, 85
\bibitem{} Dean, J. F., Warner, P. R., Cousins, A. W. J. 1978, \mnras, 183, 569
\bibitem{} Djorgovski, G. 1993, in Structure and Dynamics of Globular Clusters, 
ed. by S. G. Djorgovski and G. Meylan (ASP, San~Francisco), p. 373
\bibitem{} Djorgovski G., Piotto G., \& Capaccioli M. 1993, \aj 105, 2148
\bibitem{} Ferraro, I., Ferraro, F. R., Fusi Pecci, F., Corsi, C. E., \& 
Buonanno, R. 1995, \mnras, 275, 1057
\bibitem{} Friel, E. D. 1995, \araa, 33, 381
\bibitem{} Fusi Pecci, F., Bellazzini, M., Cacciari, C., \& Ferraro, F. R. 1995, 
\aj, 110, 1664
\bibitem{} Gnedin, O. Y., \& Ostriker, J. P. 1996, \apj, in pubblication
\bibitem{} Gratton, R. G., \& Ortolani, S. 1988, \aaps, 73, 137
\bibitem{} Grundahl, F., 1996, ASP Conf. Ser., Vol 92, 273 
\bibitem{} Hesser, J.E., Harris, W.E., VandenBerg, D.A., Allwright, J.W.B., 
Shott, P., \& Stetson, P.B. 1987, \pasp, 99, 739 (H87)
\bibitem{} Hodder, P., Nemec, J., Richer H., \& Fahlman G., 1992, \aj, 103, 460
\bibitem{} Johnstone, D. 1993, \aj, 105, 155
\bibitem{} Kaluzny J., 1994, \aaps, 108, 151 
\bibitem{} Kaluzny J., Wysocka A., Krzeminski W., 1997 (In preparation)
\bibitem{} King, I. R., 1966, \aj, 71, 64
\bibitem{} King, I. R., Hedeman, E., Hodge, S., \& White, R. 1968, \aj, 73, 456
\bibitem{} Kinman, T. D., Rosino, L. 1962, \pasp, 74, 499
\bibitem{} Landolt, A. 1992, \aj, 104, 340
\bibitem{} Lin, D. N. C., \& Richer, H. B. 1992, \apj, 388, 57
\bibitem{} Lynden-Bell, D., \& Lynden-Bell, R. M. 1995, \mnras, 275, 429
\bibitem{} Majewski, S. R. 1993, \araa, 31, 575
\bibitem{} Mandushez, G., Spassova, N. \& Staneva, A., 1991, \aap, 252, 94
\bibitem{} Ortolani, S., \& Rosino, L. 1985, Mem.S.A.It., 56, 105
\bibitem{} Picard, A., \& Johnston, P. 1993, \aap, 276, 331
\bibitem{} Piotto, G., King, I. R., Capaccioli, M., Ortolani, S., \& Djorgovski, 
G. S. 1990, \apj, 350, 662
\bibitem{} Reid M.J., 1993, \araa, 31, 345 
\bibitem{} Richer, H., \& Fahlman, G., 1988, \apj, 325, 218
\bibitem{} Rosenberg A., Piotto G., Saviane I., Aparicio A. \& Gratton R., 1997, 
ApJL (Submitted)
\bibitem{} Searle, L., \& Zinn, R., 1978, \apj, 225, 357
\bibitem{} Stetson, P. B., 1987, \pasp, 99, 191
\bibitem{} Stetson, P. B., Hesser J. E., Smith, G. H., VandenBerg, \& D. A. 
Bolte, M. 1989, \aj 96, 909
\bibitem{} Trager, S. C., King, I. R., \& Djorgovski, G. S., 1995, \aj, 109, 218
\bibitem{} VandenBerg, D. A., Bolte, M., \& Stetson, P. B. 1990, \aj, 100, 445
\bibitem{} Webbink, R. 1985, IAU Symp. Dynamics of Stars Clusters, eds. J. 
Godman and P. Hut, p. 541
\end{thebibliography}
\end{document}